\documentclass{PoS}
\pdfoutput=1
\usepackage{bm}
\newcommand{\pvec}{\bm{p}}
\newcommand{\xvec}{\bm{x}}

\newcommand{\beq}{\begin{equation}}
\newcommand{\eeq}{\end{equation}}
\newcommand{\beqy}{\begin{eqnarray}}
\newcommand{\eeqy}{\end{eqnarray}}

\newcommand{\me}[3]{\langle #1\vert\ #2\ \vert #3\rangle}
\newcommand{\ex}{\widehat{\bm{x}}}
\newcommand{\ey}{\widehat{\bm{y}}}
\newcommand{\ez}{\widehat{\bm{z}}}

\title{Spectrum of excited states using the stochastic LapH method}

\ShortTitle{Excited states with stochastic LapH}

\author{John Bulava$^a$, Brendan Fahy$^b$, Justin Foley$^c$, You-Cyuan Jhang$^b$, 
        Keisuke~J.~Juge$^{d,e}$, David~Lenkner$^b$,
        \speaker{Colin Morningstar}$^b$, and 
        Chik Him~Wong$^f$\thanks{Poster presenter.}\\
\llap{$^a$} School of Mathematics, 
            Trinity College, Dublin 2, Ireland\\
\llap{$^b$} Dept.~of Physics, Carnegie Mellon University, 
        Pittsburgh, PA 15213, USA\\
\llap{$^c$} Dept.~of Physics and Astronomy, University of Utah, 
        Salt Lake City, UT 84112, USA\\
\llap{$^d$} High Energy Accelerator Research Organization (KEK), 
        Ibaraki 305-0801, Japan\\
\llap{$^e$} Dept.~of Physics, University of the Pacific, 
        Stockton, CA 95211, USA\\
\llap{$^f$} Dept.~of Physics, University of California San Diego,
        La Jolla, CA 92093, USA}

\abstract{Progress in computing the spectrum of excited baryons and mesons 
in lattice QCD is described.  Our first results in the zero-momentum 
bosonic $I=1,\ S=0,\ T_{1u}^+$ 
symmetry sector of QCD using a correlation matrix of 56 operators are presented.
In addition to a dozen spatially-extended meson operators, 44 two-meson operators 
are used, involving a wide variety of light
isovector, isoscalar, and strange meson operators of varying relative momenta.  All needed 
Wick contractions are efficiently evaluated using a stochastic method of treating the
low-lying modes of quark propagation that exploits Laplacian Heaviside
quark-field smearing.  Level identification is discussed.}

\FullConference{31st International Symposium on Lattice Field Theory LATTICE 2013\\
		 July 29, 2013 - August 3, 2013\\
		 Mainz, Germany}

\begin{document}

\section{Introduction}

In a series of papers\cite{baryons2005A,baryons2005B,baryon2007,nucleon2009,
Bulava:2010yg,StochasticLaph,ExtendedHadrons}, we have been striving to 
compute the finite-volume stationary-state energies of QCD using Markov-chain
Monte Carlo integration of the QCD path integrals formulated on a
space-time lattice.  In this talk and accompanying poster, our progress 
towards this goal is described.  We present our first results in the zero-momentum
bosonic $I=1,\ S=0,\ T_{1u}^+$ symmetry sector of QCD using a correlation matrix of 
56 operators.  In addition to a dozen spatially-extended meson operators, an 
unprecedented number of 44 two-meson operators are used, involving a wide variety of 
light isovector, isoscalar, and strange meson operators of varying relative momenta. 
All needed Wick contractions are efficiently evaluated using a stochastic method of 
treating the low-lying modes of quark propagation that exploits Laplacian Heaviside
quark-field smearing.   Given the large number of levels extracted, level 
identification becomes a key issue.

This paper is organized as follows.  The single-hadron and two-hadron operators
that we use are briefly reviewed in Sec.~\ref{sec:ops}.  Our method of
extracting the energies and overlaps is described in Sec.~\ref{sec:extract}.
First-pass results are then presented in Sec.~\ref{sec:results}, and the
issue of level identification is confronted.  Concluding remarks are
given in Sec.~\ref{sec:conclude}.

\section{Single-hadron and multi-hadron operators}
\label{sec:ops}

The stationary-state energies in a particular symmetry sector can be extracted 
from an $N\times N$ Hermitian correlation matrix 
   $ {\cal C}_{ij}(t)
   = \langle 0\vert\, O_i(t\!+\!t_0)\, \overline{O}_j(t_0)\ \vert 0\rangle,
   $
where the $N$ operators $\overline{O}_j$ act on the vacuum to create the states 
of interest at source time $t_0$ and are accompanied by conjugate operators $O_i$ 
that can annihilate these states at a later time $t+t_0$.  
Estimates of ${\cal C}_{ij}(t)$ are obtained with the Monte Carlo method
using the stochastic LapH method\cite{StochasticLaph} which allows all needed
quark-line diagrams to be computed.  The operators that we use have been
described in detail in Refs.~\cite{baryons2005A,StochasticLaph,ExtendedHadrons},
but a brief review of our operator design is presented below.

Our hadron operators are constructed using spatially smoothened link variables 
$\widetilde{U}_j(x)$ and spatially smeared quark fields $\widetilde{\psi}(x)$.  
The spatial links are smeared using the stout-link procedure described in 
Ref.~\cite{stoutlink}.  The smeared quark field for 
each quark flavor is defined by
\begin{equation}
\widetilde{\psi}_{a\alpha}(x) =
   {\cal S}_{ab}(x,y)\ \psi_{b\alpha}(y),
\end{equation}
where $x,y$ are lattice sites, $a,b$ are color indices, and $\alpha$ is a 
Dirac spin component. We use the Laplacian Heaviside (LapH) quark-field smearing 
scheme introduced in Ref.~\cite{distillation} and defined by
\begin{equation}
{\cal S} = 
 \Theta\left(\sigma_s^2+\widetilde{\Delta}\right),
\end{equation}
where $\widetilde{\Delta}$ is the three-dimensional gauge-covariant Laplacian
defined in terms of the stout-smeared gauge field $\widetilde{U}$, and $\sigma_s$
is the smearing cutoff parameter.  More details concerning this smearing
scheme are described in Ref.~\cite{StochasticLaph}.

All of our single-hadron operators are assemblages of basic building blocks
which are gauge-covariantly-displaced, LapH-smeared quark fields:
\begin{equation}
 q^A_{a\alpha j}= D^{(j)}\widetilde{\psi}_{a\alpha}^{(A)},
 \qquad  \overline{q}^A_{a\alpha j} = \widetilde{\overline{\psi}}_{a\alpha}^{(A)}
  \gamma_4\, D^{(j)\dagger},
\label{eq:quarkdef}
\end{equation}
where $a$ is a color index, $\alpha$ is a Dirac spin component, $A$ is a quark flavor, 
$\gamma_4$ is the temporal Dirac $\gamma$-matrix, and the displacement
$D^{(j)}$ is a product of smeared link variables:
\beq
 D^{(j)}(x,x^\prime) =
 \widetilde{U}_{j_1}(x)\ \widetilde{U}_{j_2}(x\!+\!d_2)
 \ \widetilde{U}_{j_3}(x\!+\!d_3)\dots  \widetilde{U}_{j_p}(x\!+\!d_p)
  \delta_{x^\prime,x+d_{p+1}}.
\eeq
The use of $\gamma_4$ in Eq.~(\ref{eq:quarkdef}) is convenient for obtaining baryon 
correlation matrices that are Hermitian.

We can simplify our spectrum calculations as much as possible by working with 
single-hadron operators that transform irreducibly under all symmetries of a 
three-dimensional cubic 
lattice of infinite extent or finite extent with periodic boundary conditions.
The construction of irreducible representations (irreps) of $O_h^1$ begins
with the irreps of the abelian subgroup of lattice translations.  These are 
characterized by a definite three-momentum $\pvec$ as allowed by the 
periodic boundary conditions.
Each of our meson and baryon operators which creates a three-momentum $\pvec$ is a 
linear superposition of gauge-invariant quark-antiquark and three-quark elemental 
operators of the form
\beqy
 \overline{\Phi}_{\alpha\beta}^{AB}(\pvec,t)&=&
 \sum_{\bm{x}} e^{i\pvec\cdot(\xvec+\frac{1}{2}(\bm{d}_\alpha+\bm{d}_\beta))}
   \delta_{ab}\ \overline{q}^B_{b\beta}(\bm{x},t)\ q^A_{a\alpha}(\bm{x},t),\\
  \overline{\Phi}_{\alpha\beta\gamma}^{ABC}(\pvec,t)&= &
 \sum_{\bm{x}} e^{i\pvec\cdot\xvec}\varepsilon_{abc}
\ \overline{q}^C_{c\gamma}(\bm{x},t)
\ \overline{q}^B_{b\beta}(\bm{x},t)
\ \overline{q}^A_{a\alpha}(\bm{x},t),
\eeqy
where $q,\overline{q}$ are defined in Eq.~(\ref{eq:quarkdef}),
$\bm{d}_\alpha, \bm{d}_\beta$ are the spatial displacements of the
$\overline{q},q$ fields, respectively, from $\xvec$,
$A,B$ indicate flavor, and $\alpha,\beta$ are compound indices
incorporating both spin and quark-displacement types. 
The phase factor
involving the quark-antiquark displacements is needed to ensure proper
transformation properties under $G$-parity for arbitrary displacement types.
Group theory projections onto the irreps of the lattice symmetry
group are then employed.
Each meson and baryon source operator ends up having the form
\beq
   \overline{M}_l(t)=c_{\alpha\beta}^{(l)\ast}\ \overline{\Phi}_{\alpha\beta}^{AB}(t),
  \qquad
   \overline{B}_l(t)=c_{\alpha\beta\gamma}^{(l)\ast}
   \ \overline{\Phi}_{\alpha\beta\gamma}^{ABC}(t),
\eeq
(or is a flavor combination of the above form),
where $l$ is a compound index comprised of a three-momentum $\pvec$, an
irreducible representation $\Lambda$ of the little group of $\pvec$,
the row $\lambda$ of the irrep, total isospin $I$, isospin projection $I_3$,
strangeness $S$, and an identifier labeling the different operators in each 
symmetry channel.  Here, we focus on mesons containing only $u,d,s$ quarks.
In order to build up the necessary orbital and radial structures expected
in the hadron excitations, we use a variety of spatially-extended configurations
for our hadron operators, as shown in Table~\ref{tab:opforms}.

\begin{table}
\caption[captab]{
The spatial arrangements of the quark-antiquark meson operators 
and the three-quark baryon operators. In the illustrations, the smeared quarks 
fields are depicted by solid circles, each hollow circle indicates a smeared ``barred'' 
antiquark field, the solid line segments indicate covariant displacements, and each 
hollow box indicates the location of a Levi-Civita color coupling.  For simplicity, 
all displacements have the same length in an operator. 
\label{tab:opforms}}
\begin{tabular}{c}
\raisebox{3mm}{
\begin{minipage}{\textwidth} 
\raisebox{0mm}{\setlength{\unitlength}{1mm}
\thicklines
\begin{picture}(20,12)
\put(9,7){\circle{2}}
\put(11,7){\circle*{2.5}}
\put(3,2){single-site}
\end{picture}}\hspace*{3mm}
\raisebox{0mm}{\setlength{\unitlength}{1mm}
\thicklines
\begin{picture}(20,12)
\put(7,7){\circle{2}}
\put(13,7){\circle*{2.5}}
\put(8,7){\line(1,0){4}}
\put(-1,2){singly-displaced}
\end{picture}} \hspace*{8mm}
\raisebox{0mm}{\setlength{\unitlength}{1mm}
\thicklines
\begin{picture}(20,12)
\put(7,11){\circle{2}}
\put(13,5){\circle*{2.5}}
\put(12,5){\line(-1,0){5}}
\put(7,10){\line(0,-1){5}}
\put(-3,0){doubly-displaced-L}
\end{picture}} \hspace*{10mm}
\raisebox{0mm}{\setlength{\unitlength}{1mm}
\thicklines
\begin{picture}(20,12)
\put(8,11){\circle{2}}
\put(14,11){\circle*{2.5}}
\put(8,4){\line(1,0){6}}
\put(14,4){\line(0,1){6}}
\put(8,4){\line(0,1){6}}
\put(-2,0){triply-displaced-U}
\end{picture}}\hspace*{10mm}
\raisebox{0mm}{\setlength{\unitlength}{1mm}
\thicklines
\begin{picture}(20,15)
\put(7,5){\circle{2}}
\put(11,7){\circle*{2.5}}
\put(7,12){\line(1,0){8}}
\put(7,6){\line(0,1){6}}
\put(15,12){\line(-3,-4){3.0}}
\put(-3,0){triply-displaced-O}
\end{picture}}
\end{minipage}} \\[4mm]
\raisebox{-2mm}{
\begin{minipage}{\textwidth}
\raisebox{0mm}{\setlength{\unitlength}{1mm}
\thicklines
\begin{picture}(16,10)
\put(5.25,4){\line(1,0){5.25}}
\put(5.25,9.5){\line(1,0){5.25}}
\put(5.25,4){\line(0,1){5.5}}
\put(10.5,4){\line(0,1){5.5}}
\put(7,6){\circle*{2}}
\put(9,6){\circle*{2}}
\put(8,8){\circle*{2}}
\put(1,0){single-site}
\end{picture}} \hspace*{0mm}
\raisebox{0mm}{\setlength{\unitlength}{1mm}
\thicklines
\begin{picture}(23,10)
\put(5.25,3.5){\line(1,0){3.5}}
\put(5.25,3.5){\line(0,1){5.5}}
\put(5.25,9.0){\line(1,0){3.5}}
\put(8.75,3.5){\line(0,1){5.5}}
\put(7,5){\circle*{2}}
\put(7,7.3){\circle*{2}}
\put(14,6){\circle*{2}}
\put(8.75,6){\line(1,0){4.5}}
\put(4,0){singly-}
\put(2,-4){displaced}
\end{picture}}  \hspace*{0mm}
\raisebox{0mm}{\setlength{\unitlength}{1mm}
\thicklines
\begin{picture}(26,8)
\put(10.4,3.5){\line(1,0){3}}
\put(10.4,3.5){\line(0,1){3}}
\put(10.4,6.5){\line(1,0){3}}
\put(13.4,3.5){\line(0,1){3}}
\put(12,5){\circle*{2}}
\put(6,5){\circle*{2}}
\put(18,5){\circle*{2}}
\put(6,5){\line(1,0){4.2}}
\put(18,5){\line(-1,0){4.4}}
\put(5,0){doubly-}
\put(3,-4){displaced-I}
\end{picture}} \hspace*{0mm}
\raisebox{0mm}{\setlength{\unitlength}{1mm}
\thicklines
\begin{picture}(20,13)
\put(6.5,3.5){\line(1,0){3}}
\put(6.5,3.5){\line(0,1){3}}
\put(6.5,6.5){\line(1,0){3}}
\put(9.5,3.5){\line(0,1){3}}
\put(8,5){\circle*{2}}
\put(8,11){\circle*{2}}
\put(14,5){\circle*{2}}
\put(14,5){\line(-1,0){4.4}}
\put(8,11){\line(0,-1){4.2}}
\put(4,0){doubly-}
\put(1,-4){displaced-L}
\end{picture}}  \hspace*{0mm}
\raisebox{0mm}{\setlength{\unitlength}{1mm}
\thicklines
\begin{picture}(20,12)
\put(9,9){\line(1,0){2}}
\put(9,11){\line(1,0){2}}
\put(9,9){\line(0,1){2}}
\put(11,9){\line(0,1){2}}
\put(4,10){\circle*{2}}
\put(16,10){\circle*{2}}
\put(10,4){\circle*{2}}
\put(4,10){\line(1,0){5}}
\put(16,10){\line(-1,0){5}}
\put(10,4){\line(0,1){5}}
\put(5,0){triply-}
\put(1,-4){displaced-T}
\end{picture}}   \hspace*{0mm}
\raisebox{0mm}{\setlength{\unitlength}{1mm}
\thicklines
\begin{picture}(20,12)
\put(9,9){\line(1,0){2}}
\put(9,11){\line(1,0){2}}
\put(9,9){\line(0,1){2}}
\put(11,9){\line(0,1){2}}
\put(6,6){\circle*{2}}
\put(16,10){\circle*{2}}
\put(10,4){\circle*{2}}
\put(9,9){\line(-1,-1){3.6}}
\put(16,10){\line(-1,0){5}}
\put(10,4){\line(0,1){5}}
\put(5,0){triply-}
\put(1,-4){displaced-O}
\end{picture}} \vspace*{3mm}
\end{minipage} }
\end{tabular}
\end{table}

We construct our two-hadron operators as superpositions of single-hadron 
operators of definite momenta.
Each single-hadron operator is labelled by total isospin $I$, the projection
of the total isospin $I_3$, strangeness $S$, three-momentum $\pvec$,
the little group irrep $\Lambda$, the row of the irrep $\lambda$,
and $i$, which denotes all other identifying information, such as
the displacement type and index.  Hence, the two-hadron operators 
have the form
\beq
    c^{I_{3a}I_{3b}}_{\pvec_a\lambda_a;\ \pvec_b\lambda_b}
     \ B^{I_aI_{3a}S_a}_{\pvec_a\Lambda_a\lambda_a i_a}
     \  B^{I_bI_{3b}S_b}_{\pvec_b\Lambda_b\lambda_b i_b},
\eeq
for fixed total momentum $\pvec=\pvec_a+\pvec_b$, and fixed 
$\Lambda_a, i_a, \Lambda_b, i_b$. Once again, group theory projections 
onto the little group of $\pvec$ and the isospin irreps are carried out.
For practical reasons, we restrict our attention to certain classes of 
momentum directions for the single hadron operators: on axis 
$\pm\ex,\ \pm\ey,\ \pm\ez$, planar diagonal 
$\pm\ex\pm\ey,\ \pm\ex\pm\ez,\ \pm\ey\pm\ez$,
and cubic diagonal $\pm\ex\pm\ey\pm\ez$.
It is crucial to know and fix all phases of the single-hadron 
operators for all momenta.  To do this, we choose a reference
direction $\pvec_{\rm ref}$ for each class of momentum directions,
and for each $\pvec$, we select one reference rotation 
$R_{\rm ref}^{\pvec}$ that transforms $\pvec_{\rm ref}$ into $\pvec$.
The details are described in Ref.~\cite{ExtendedHadrons}.
This approach is efficient for creating large numbers of two-hadron 
operators, and generalizes to three or more hadrons.

\begin{figure}
\begin{center}
\includegraphics[trim=0mm 20mm 0mm 80mm,width=1.8in,clip=true]{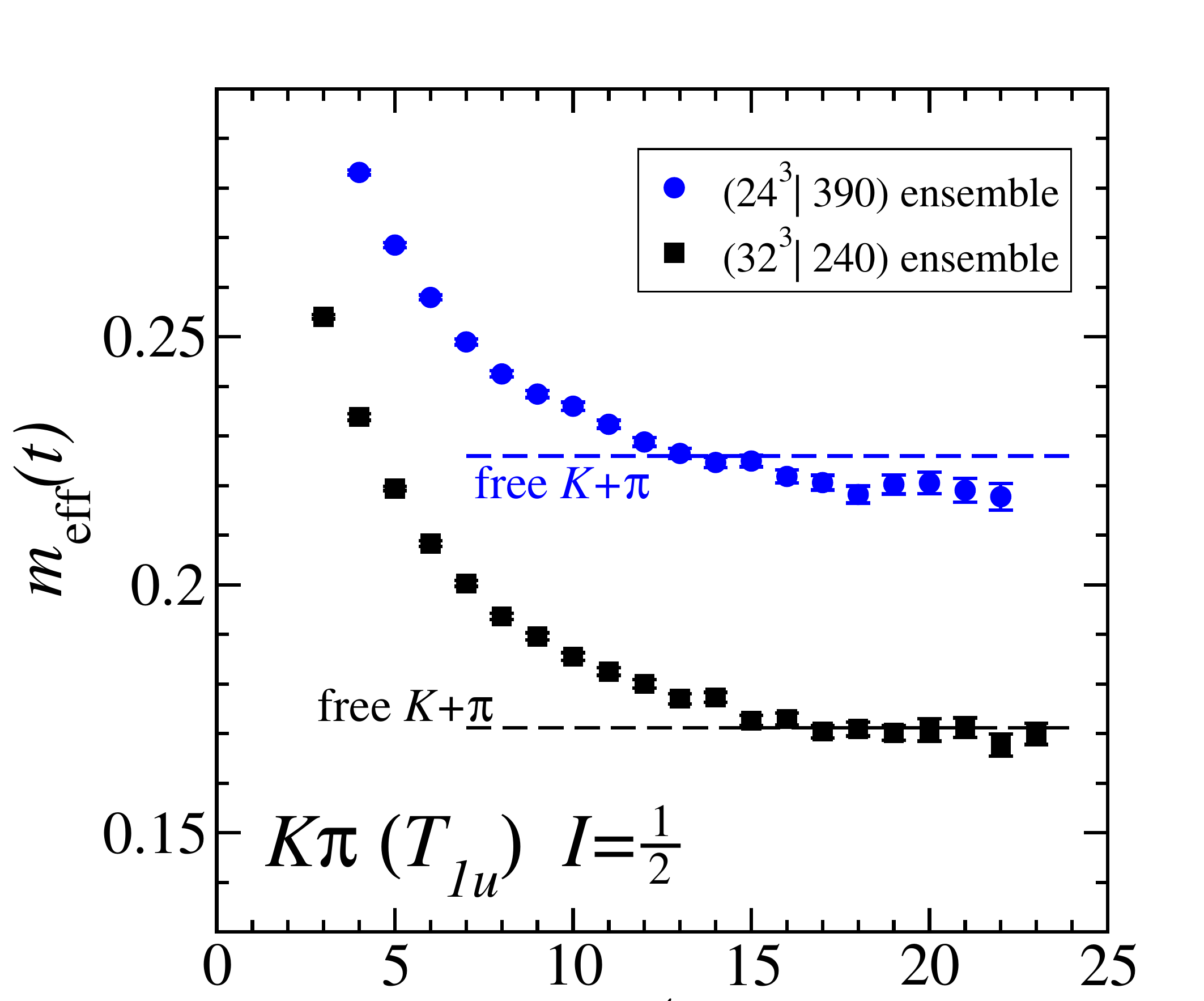}
\includegraphics[trim=0mm 20mm 0mm 80mm,width=1.8in,clip=true]{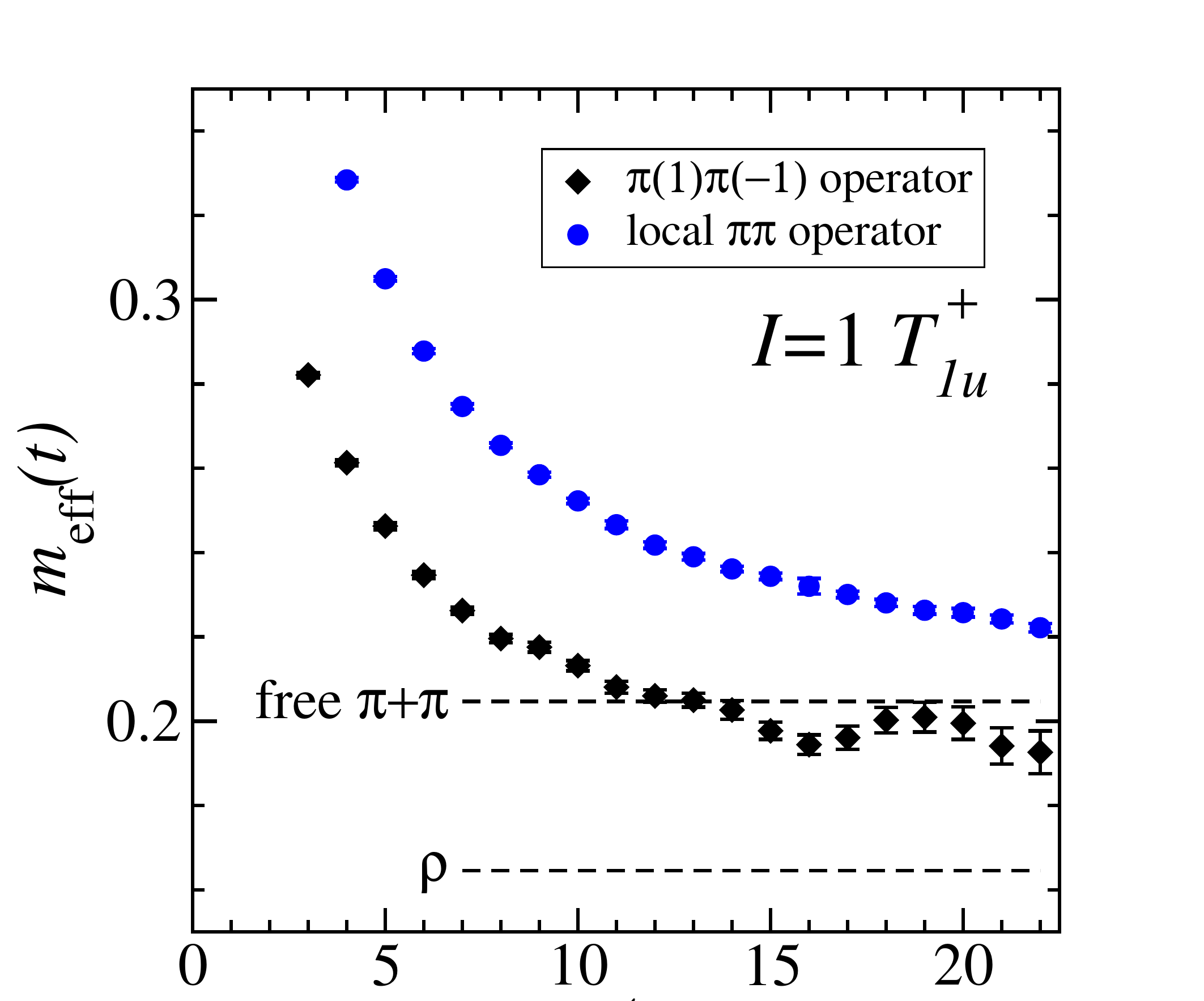}
\includegraphics[trim=0mm 20mm 0mm 80mm,width=1.8in,clip=true]{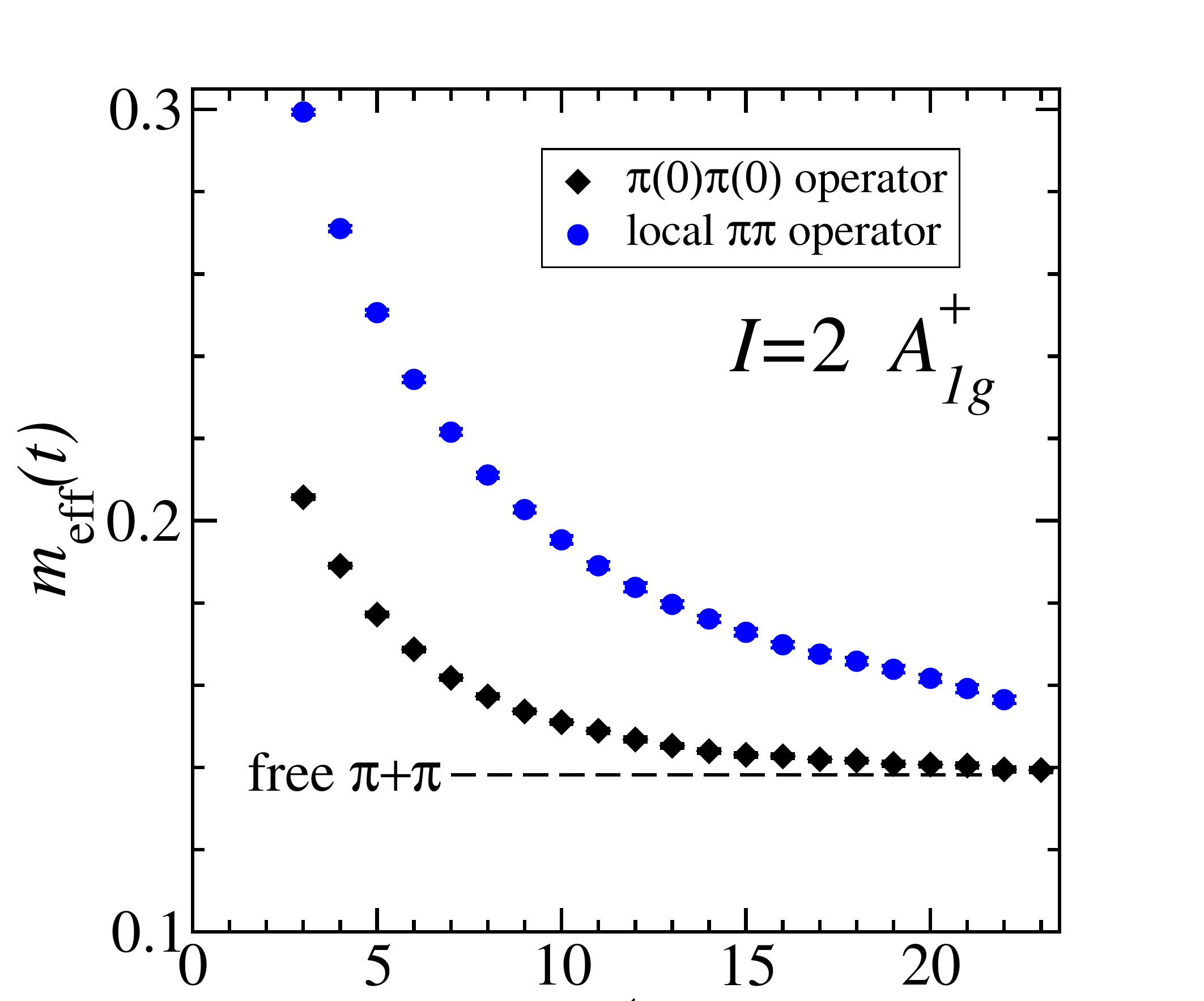}
\end{center}
\caption[kpi]{
(Left) Effective masses, $m_{\rm eff}(t)$, associated with a 
two-meson operator in the $T_{1u}$ irrep, having total isospin $I=\frac{1}{2}$ and 
zero total momentum, constructed from single-site kaon and pion operators having 
equal and opposite on-axis momenta of minimal nonzero magnitude.  Results on the 
$(24^3\vert 390)$ and $(32^3\vert 240)$ ensembles (see text) are shown.  The energies
of a free $\pi$ plus a free $K$ are indicated by horizontal dashed lines.  
(Center) Effective mass for one of our $I=1$ $\pi(1)\pi(-1)$ operators in the 
$T_{1u}^+$ channel, consisting of single-site pion operators having equal and 
opposite on-axis momenta of minimal nonzero magnitude, compared to the effective 
mass of a localized $\pi\pi$ operator, described in Eq.~(\ref{eq:localpipiT1um}),
on the $(24^3\vert 390)$ ensemble.  (Right) Effective mass for one of our 
$I=2$ $\pi(0)\pi(0)$ operators in the $A_{1g}^+$ channel, consisting of single-site 
pion operators each having zero momenta, compared to the effective mass of a 
localized $\pi\pi$ operator, described in Eq.~(\ref{eq:localpipiA1gp}), on 
the $(24^3\vert 390)$ ensemble. 
\label{fig:kaonpion}}
\end{figure}

In addition to efficiency, there are good physical reasons for using
such multi-hadron operators.  Hadron-hadron interactions in finite volume
move the energies of any two-hadron systems away from their 
free two-particle energies, and the interacting two-particle states
could involve distributions of different relative momenta.  
However, such interactions are usually small and the relative momenta
used in our operators should presumably dominate in most cases.  
Also, we will always utilize multi-hadron operators with a variety of
different relative momenta to accommodate the effects of such interactions.
The performance of some of our $\pi\pi$ operators are compared to
localized multi-hadron operators in Fig.~\ref{fig:kaonpion}, discussed below.

In order to test the effectiveness of the two-hadron operators
that we have designed, we examined the effective masses associated with
the correlators of a variety of two-hadron operators.  
In the left plot of Fig.~\ref{fig:kaonpion}, effective masses 
associated with a two-meson operator in the $T_{1u}$ irrep are shown.  
The two-meson operator has total isospin $I=\frac{1}{2}$ and zero total 
momentum and is constructed from single-site kaon and pion operators having 
equal and opposite on-axis momenta of minimal nonzero magnitude.  Results 
on the $(24^3\vert 390)$ and $(32^3\vert 240)$ ensembles (described below)
are shown and compared to the energies of a free $\pi$ plus a free $K$, 
indicated by horizontal dashed lines.

An alternative design for a two-hadron operator is to use a suitable
localized-field operator.  For example, localized $\pi\pi$ operators in the
$I=2,\ A_{1g}^+$ and $I=1,\ T_{1u}^+$ channels can be obtained using
\begin{eqnarray}
 (\pi\pi)^{A_{1g}^+}(t) &=& \sum_{\bm{x}}
  \pi^+(\bm{x},t)\ \pi^+(\bm{x},t),
 \label{eq:localpipiA1gp}\\
 (\pi\pi)^{T_{1u}^+}(t) &=& \!\!\!\!\!\sum_{\bm{x},k=1,2,3}
 \!\!\!\!\Bigl\{ \pi^+(\bm{x},t)\ \Delta_k\pi^0(\bm{x},t) 
 -\pi^0(\bm{x},t)\ \Delta_k\pi^+(\bm{x},t)\Bigr\},
 \label{eq:localpipiT1um}
\end{eqnarray}
where $\pi(\bm{x},t)$ is a single-site pion field using a standard
$\gamma_5$ construction with the LapH-smeared quark fields, and
$\Delta_k\pi(\bm{x},t)=\pi(\bm{x}\!+\!\widehat{\bm{k}},t)
 -\pi(\bm{x}\!-\!\widehat{\bm{k}},t)$.  The superscripts indicate
the electric charges associated with each field. In such localized $\pi\pi$
operators, the individual pions do not have definite momenta.

The center and right plots of Fig.~\ref{fig:kaonpion} compare the
effective masses for our $\pi\pi$ operators to those for these
localized $\pi\pi$ operators.  The center plot of Fig.~\ref{fig:kaonpion}
shows the effective mass for one of our $I=1$ $\pi(1)\pi(-1)$ 
operators in the $T_{1u}^+$ channel, consisting of single-site pion operators 
having equal and opposite on-axis momenta of minimal nonzero magnitude, 
compared to the effective mass of the localized $\pi\pi$ operator, given 
in Eq.~(\ref{eq:localpipiT1um}), on the $(24^3\vert 390)$ ensemble.  The
right plot of Fig.~\ref{fig:kaonpion} shows the effective mass for one of our 
$I=2$ $\pi(0)\pi(0)$ operators in the $A_{1g}^+$ channel, consisting of 
single-site pion operators each having zero momenta, compared to the 
effective mass of the localized $\pi\pi$ operator, given in 
Eq.~(\ref{eq:localpipiA1gp}), on the $(24^3\vert 390)$ ensemble. 
One sees that the effective masses of the localized $\pi\pi$ operators
lie well above those of our operators, indicating that they contain much 
more excited-state contamination.  These effective masses are compared
to the energies of the ground state $\rho$ and the free $\pi+\pi$ energies,
indicated by horizontal dashed lines, in this figure.  Note that, in
addition to having much less excited-state contamination, the two-pion 
operators comprised of individual pions having definite momenta are
also much easier to make in large numbers, compared to the localized
multi-hadron operators.

\section{Energy and overlap extractions}
\label{sec:extract}

For the large numbers of operators we plan to use, it can happen that
the condition number of the correlation matrix can grow uncomfortably
large.  In a large matrix of correlations, statistical noise can cause the 
matrix to become ill-conditioned, or even to have negative eigenvalues so
that the matrix is no longer positive definite.  It is important to monitor 
this and take corrective actions.  We do this as follows.  
Starting with a raw correlation matrix ${\cal C}(t)$, we first try to remove the 
effects of differing normalizations by forming the matrix 
$ \widehat{C}_{ij}(t)={\cal C}_{ij}(t)\ (\ {\cal C}_{ii}(\tau_N){\cal C}_{jj}(\tau_N)\ )^{-1/2}$,
taking $\tau_N$ at a very early time, such as $\tau_N=3$.  
We then pick a large value $t=t_F$ and evaluate the
eigenvalues and eigenvectors of $\widehat{C}(t_F)$. Since the matrix is Hermitian, 
the eigenvalues are real and the eigenvectors are orthonormal.  Let $U_N$ 
denote the unitary matrix whose columns are the eigenvectors of $\widehat{C}(t_F)$.  
The columns corresponding to negative eigenvalues must be removed.  We also 
remove the columns corresponding to eigenvalues that are positive, but small. 
In other words, we remove the columns from $U_N$ for all eigenvalues less 
than some threshold $\lambda_{\rm threshold}$.  Let $P_N$ denote the
projection matrix whose columns are the retained columns (eigenstates)
of $U_N$.  We then apply this projection to the correlators to obtain 
a new correlation matrix $C(t)$:
\beq
     C(t)=   P_N^\dagger\ \widehat{C}(t)\ P_N.
\eeq
The threshold $\lambda_{\rm threshold}$ is determined as follows.
We decide on the largest value of the condition number that is
acceptable, denoting this by $\xi^{\rm cn}_{\rm max}$.
We determine the largest eigenvalue $\lambda_{\rm max}$, then since
the condition number is the ratio of the largest eigenvalue over the
eigenvalue of smallest magnitude, the minimum allowed eigenvalue
is
\beq
     \lambda_{\rm threshold} = \frac{\lambda_{\rm max}}{
   \xi^{\rm cn}_{\rm max}}.
\eeq
The above procedure ensures that $C(t_F)$ is positive-definite
and well conditioned.  We also check this for all other $t$ values
that we use.  Extraction of the energies then proceeds using the
refined correlator $C(t)$.  

In finite volume, all energies are discrete so that each correlator matrix
element has a spectral representation of the form
\beq
   C_{ij}(t) = \sum_n Z_i^{(n)} Z_j^{(n)\ast}\ e^{-E_n t},
   \qquad\quad Z_j^{(n)}=  \me{0}{O_j}{n},
\eeq
assuming temporal wrap-around (thermal) effects are negligible.
For temporal separations $t$ large enough such that only the lowest $N$ energies 
contribute, we can solve for $E_n$ and $ Z_j^{(n)}$ using the correlation matrix
at two time separations $C(\tau_0)$ and $C(t)$.  We first
solve the generalized eigenvector problem $Ax=\lambda Bx$ with
$A=C(t)$ and $B=C(\tau_0)$.  We put the eigenvectors into the columns
of the matrix $V$ with normalization condition $V^\dagger C(\tau_0)V=I$,
and the eigenvalue corresponding to column $n$ is
$\lambda_n=e^{-E_n(t-\tau_0)}$.  Then the
overlaps are given by
\[
    Z_{j}^{(n)} = C_{jk}(\tau_0)\ V_{kn}\ e^{E_n\tau_0/2}.
\]
If $\tau_0$ and $t$ are chosen such that the contributions to all 
$C_{ij}(\tau_0)$ and $C_{ij}(t)$ from \textit{all} levels above the 
lowest $N$ energies are negligible, then the results obtained for the 
$E_n$ and $Z_{j}^{(n)}$ will be the same for all suitably large
$\tau_0$ and $t$.  To check this, one usually fixes $\tau_0$
and varies $t$, making $t$ larger and larger until the solved
values for $E_n$ and $Z_{j}^{(n)}$ become independent of $t$.

For fixed $\tau_0$ (which must be suitably large), these solutions for 
$E_n$ and $Z_{j}^{(n)}$ can be viewed as functions of $t$, and one can
even perform fits to these functions with simple empirically-motivated forms
to extract their large $t$ values.
For example, the $N$ eigenvalues of $C(\tau_0)^{-1/2}\ C(t)\ C(\tau_0)^{-1/2}$
are known as the \textit{principal correlators} and are denoted by
$\lambda_\alpha(t,\tau_0)$.  As $t$ becomes large,
\beq
 \lim_{t\rightarrow \infty}
    \lambda_\alpha(t,\tau_0)= e^{-(t-\tau_0)E_\alpha}.
\eeq
This already complicated procedure is further plagued by the issue of
eigenvector identification. Eigenvector ``pinning'' is usually needed in this 
method to deal with level switching. Since the eigenvalue equation is solved 
for eigenvalues $E_n$ and eigenvectors $v^{(n)}$, independently on each 
timeslice $t$ and for each bootstrap sample, ensuring the same ordering of 
states between time slices and bootstrap samples requires some care,
especially when levels are closely spaced or nearly degenerate.
Instead of ordering by the value of the eigenvalue, one associates states 
between neighboring time slices and bootstrap samples using the similarity
of their eigenvectors.  For ordering the states on time $t$ using the full
ensemble, one uses the eigenvectors obtained on the full ensemble (not a 
bootstrap sample) on timeslice $t-1$ as the reference eigenvectors 
$v^{(n)}_{\rm ref}$.  The eigenvector comparison is done by finding
the maximum value of $v^{(n^\prime)\dagger}_{\rm ref}C(\tau_0)v^{(n)}$
which associates a state $n$ with a reference state $n^\prime$. 
When ordering levels in a bootstrap sample for time $t$, one uses
the eigenvectors obtained on the full ensemble for time $t$ as
the references.

The above ``principal correlator method'' is rather costly and complicated.
Diagonalizations must be carried out for a large number of times $t$, and
the diagonalizations at large times can amplify errors and possibly introduce
bias.  Also, the eigenvector ``pinning'' is somewhat ad hoc, and hence,
unpalatable. A simpler method, which here is called the ``single rotation'' 
method, or the ``fixed coefficient'' method, performs the diagonalization with 
one choice of metric time $\tau_0$ and one time $t=\tau_D$.  The eigenvectors 
obtained are used to ``rotate'' the original correlator $C(t)$ into a correlator 
$G(t)$ for which $G(\tau_0)=1$, the identity matrix, 
and $G(\tau_D)$ is diagonal.  At other times, $G(t)$ 
need not be diagonal.  However, with judicious choices of $\tau_0$ and $\tau_D$, 
one finds that the off-diagonal elements of $G(t)$ remain zero within 
statistical precision 
for $t>\tau_D$. The rotated correlator is given by
\beq
 G(t) = U^\dagger\ C(\tau_0)^{-1/2}\ C(t)\ C(\tau_0)^{-1/2}\ U,
\label{eq:rotatedcorr}
\eeq
where the columns of $U$ are the orthonormalized eigenvectors of
$C(\tau_0)^{-1/2}\ C(\tau_D)\ C(\tau_0)^{-1/2}$.
Rotated effective masses can then be defined by
\beq
  m_G^{(n)}(t)=\frac{1}{\Delta t}
  \ln\left(\frac{G_{nn}(t)}{
  G_{nn}(t+\Delta t)}\right),
\label{eq:roteffmass}
\eeq
which tend to the lowest-lying $N$ stationary-state energies
produced by the $N$ operators.  Correlated-$\chi^2$ fits to 
the estimates of $G_{nn}(t)$ using the forms $A_n(e^{-E_n\,t}+e^{-E_n\,(T-t)})$,
where $T$ is the temporal extent of the lattice, yield the energies $E_n$
and the overlaps $A_n$ to the rotated operators for each $n$. Using the 
rotation coefficients, one can then easily obtain the overlaps 
$Z^{(n)}_j=C(\tau_0)^{1/2}_{jk}\ U_{kn}\ A_n$ (no summation over $n$)
corresponding to the rows and columns of the correlation matrix $C(t)$.
Use of the $P_N$ matrix gives the overlaps for the original operator set,
modulo a normalization factor for each operator.

\section{First results}
\label{sec:results}

We are currently focusing on three Monte Carlo ensembles: (A) a set of 
412 gauge-field configurations on a large $32^3\times 256$ anisotropic lattice 
with a pion mass $m_\pi\sim 240$~MeV, (B) an ensemble of 551 configurations
on an $24^3\times 128$ anisotropic lattice with a pion mass
$m_\pi\sim 390$~MeV, and (C) an ensemble of 584 configurations
on an $24^3\times 128$ anisotropic lattice with a pion mass
$m_\pi\sim 240$~MeV.  We refer to these ensembles as the 
$(32^3\vert 240)$, $(24^3\vert 390)$, and $(24^3\vert 240)$ ensembles,
respectively.  These ensembles were generated using the Rational 
Hybrid Monte Carlo (RHMC) algorithm\cite{Clark:2004cp}. In each ensemble, successive 
configurations are separated by 20 RHMC trajectories to minimize autocorrelations.
An improved anisotropic clover fermion action and an improved gauge field 
action are used\cite{Lin:2008pr}.  In these ensembles, $\beta=1.5$
and the $s$ quark mass parameter is set to $m_s=-0.0743$ in order to reproduce 
a specific combination of hadron masses\cite{Lin:2008pr}.  
In the $(24^3\vert 390)$ ensemble, the light quark mass parameters are set to
$m_u=m_d=-0.0840$ so that the pion mass is around 390~MeV if one sets the scale 
using the $\Omega$ baryon mass.   In the $(32^3\vert 240)$ and 
$(24^3\vert 240)$ ensembles, $m_u=m_d=-0.0860$ are used, resulting in a pion 
mass around 240~MeV.  The spatial grid size is $a_s\sim 0.12$~fm, whereas
the temporal spacing is $a_t\sim 0.035$~fm.

In our operators, a stout-link staple weight $\xi=0.10$ is used with
$n_\xi=10$ iterations.  For the cutoff in the LapH smearing, we use 
$\sigma_s^2=0.33$,  which translates into the number $N_v$ of LapH eigenvectors 
retained being $N_v=112$ for the $24^3$ lattices and $N_v=264$ for the $32^3$ 
lattice. We use $Z_4$ noise in all of our stochastic estimates of quark propagation.
Our variance reduction procedure is described in Ref.~\cite{StochasticLaph}.
On the $24^3$ lattices, we use 4 widely-separated source times $t_0$,
and 8 are used on the $32^3$ lattice.

\begin{figure}
\includegraphics[width=5.8in]{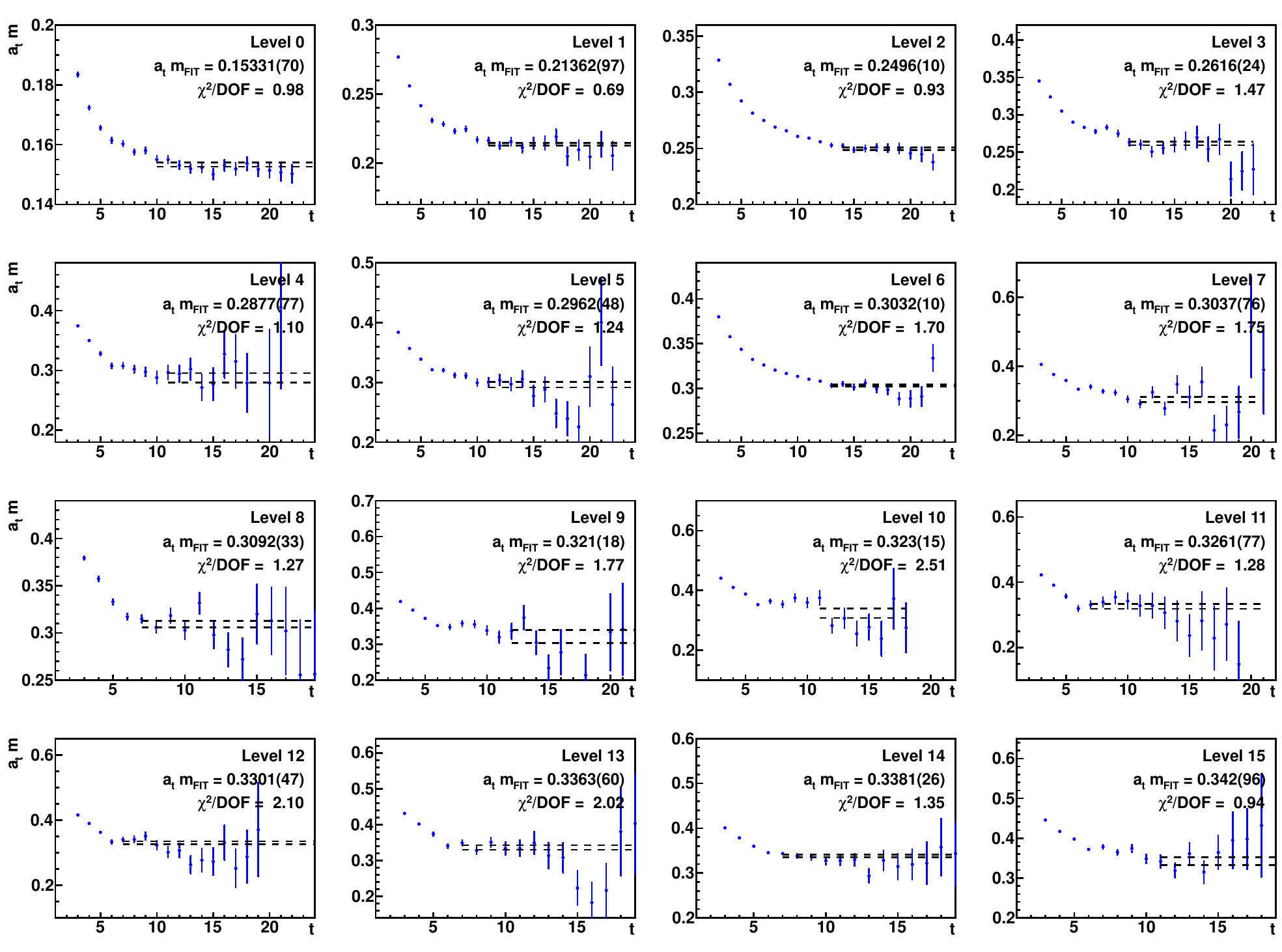}
\caption[cap1]{
Rotated effective masses $m_G^{(n)}(t)$ 
(see Eq.~(\ref{eq:roteffmass})) for the 16 lowest-lying energy levels in the 
zero-momentum bosonic $I=1,\ S=0,\ T_{1u}^+$
channel for the $(24^3\vert 390)$ ensemble using 12 single-meson
operators, 17 isovector+isovector operators, 17 isoscalar+isovector
operators, and 10 kaon+antikaon operators.  Dashed lines indicate
energy extractions from correlated-$\chi^2$ fits.  Fit results
and qualitites are also listed in each plot.
\label{fig:levels1}}
\end{figure}  

The Monte Carlo method commonly employed in QCD computations applies only to
space-time lattices of finite extent.  Hence, the energies we extract are
those associated with the stationary states of QCD in a cubic box using periodic 
boundary conditions.  In such a cubic box, we no longer have full rotational 
symmetry, even in the continuous space-time limit.
The stationary states cannot be labelled by the usual spin-$J$ quantum
numbers. Instead, the stationary states in a box with periodic boundary 
conditions must be labelled by the irreducible representations (irreps) of 
the cubic space group, even in the continuum limit.  A detailed description
of these irreps is summarized in Ref.~\cite{ExtendedHadrons}.

\begin{figure}
\includegraphics[width=5.8in]{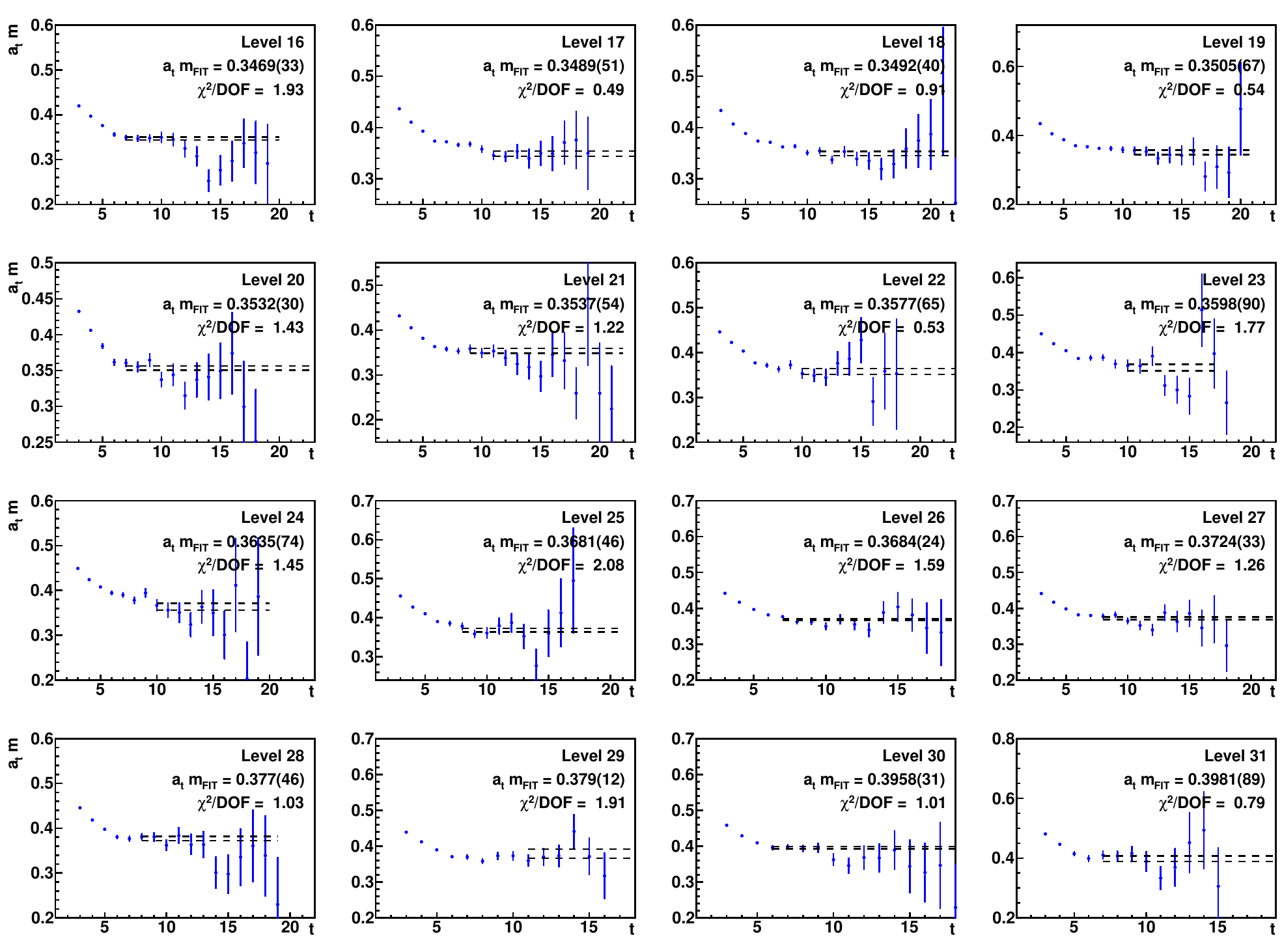}
\caption[cap2]{
Rotated effective masses $m_G^{(n)}(t)$ 
(see Eq.~(\ref{eq:roteffmass})) for energy levels 16 to 31 in the 
zero-momentum bosonic $I=1,\ S=0,\ T_{1u}^+$
channel for the $(24^3\vert 390)$ ensemble using 12 single-meson
operators, 17 isovector+isovector operators, 17 isoscalar+isovector
operators, and 10 kaon+antikaon operators.  Dashed lines indicate
energy extractions from correlated-$\chi^2$ fits.  Fit results
and qualitites are also listed in each plot.
\label{fig:levels2}}
\end{figure}  

\begin{figure}
\includegraphics[width=5.8in]{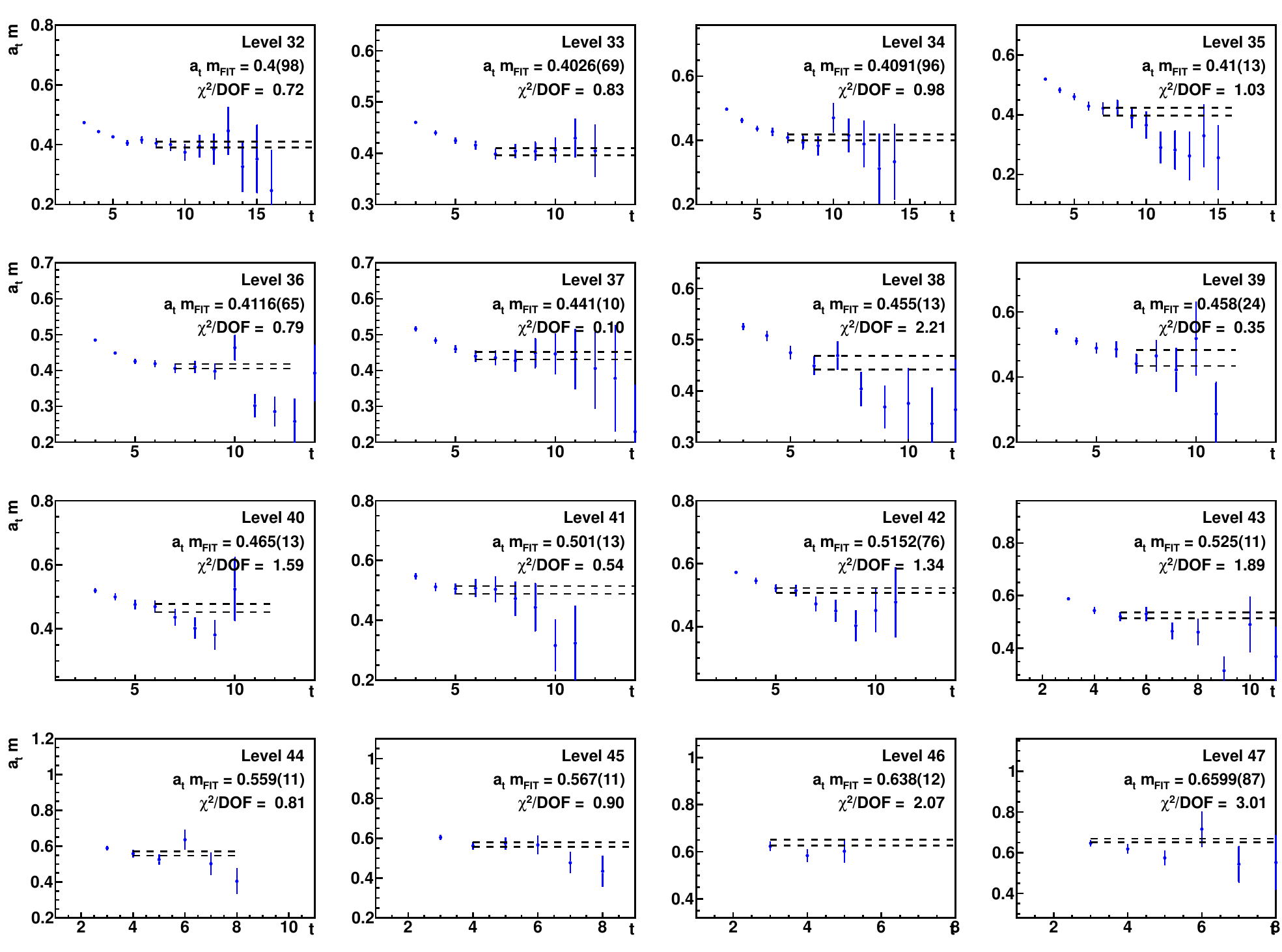}
\caption[cap3]{
Rotated effective masses $m_G^{(n)}(t)$ 
(see Eq.~(\ref{eq:roteffmass})) for energy levels 32 to 47 in the 
zero-momentum bosonic $I=1,\ S=0,\ T_{1u}^+$
channel for the $(24^3\vert 390)$ ensemble using 12 single-meson
operators, 17 isovector+isovector operators, 17 isoscalar+isovector
operators, and 10 kaon+antikaon operators.  Dashed lines indicate
energy extractions from correlated-$\chi^2$ fits.  Fit results
and qualitites are also listed in each plot.
\label{fig:levels3}}
\end{figure}  

For our first calculations, we decided to focus on the resonance-rich
$I=1,\ S=0,\ T_{1u}^+$ channel of total zero momentum.  This channel has 
odd parity, even $G$-parity, and contains the spin-1 and spin-3 mesons.
The exper\-imentally-known resonances in this
channel include the $\rho(770)$, $\rho(1450)$, $\rho(1570),$ $\rho_3(1690),$
and $\rho(1700)$.  Low statistics runs on smaller lattices led us to
include 12 particular single-meson (quark-antiquark) operators.
We took special care to include operators that could produce the 
spin-$3$ $\rho_3(1690)$ state, in addition to the other spin-$1$ states.
Low statistics runs also gave us the masses of the lowest-lying
mesons, such as the $\pi,\eta, K,$ and so on.  Given these known
mesons, we used software written in \textsc{Maple} to find all possible
two-meson states in our cubic box in this $T_{1u}^+$ symmetry
channel, assuming no energy shifts from interactions or the
finite volume.  We used these so-called ``expected two-meson levels''
to guide our choice of two-meson operators to include.  We
included 17 isovector-isovector meson operators, 14 operators
that combine an isovector with a light isoscalar (using only $u,d$
quarks), 3 operators that combine an isovector with an
$\overline{s}s$ isoscalar meson, and 10 kaon-antikaon operators.
Our ``first-pass'' results for the $(24^3\vert 390)$ ensemble
obtained from our $56\times 56$ correlation matrix
are presented in 
Figs.~\ref{fig:levels1}, \ref{fig:levels2}, and \ref{fig:levels3}.
The rotated effective masses $m_G^{(n)}(t)$ 
(see Eq.~(\ref{eq:roteffmass})) using $\tau_0=5$ and
$\tau_D=9$ are shown in these figures.  Fits values and qualities
are listed, and depicted by the horizontal dashed lines.

The results shown here are not finalized yet.  We are still
varying the fitting ranges to improve the $\chi^2$, as needed in
some instances.  We are investigating the effects of adding more
operators, and we are even still verifying our analysis/fitting
software.  However, these figures do demonstrate that the extraction
of a large number of energy levels is indeed possible, and the
plots indicate the level of precision that can be attained with
our stochastic LapH method.  Keep in mind that we have not included 
any three-meson operators in our correlation matrix. 

With such a large number of energies extracted, level identification 
becomes a key issue.  QCD is a complicated interacting quantum field
theory, so characterizing its stationary states in finite volume
is not likely to be done in a simple way.  Level identification must
be inferred from the $Z$ overlaps of our probe operators, analogous
to deducing resonance properties from scattering cross sections
in experiments.  Although we are in control of the probe operators
$\overline{O}_j$ which act on the vacuum to create ``probe
states'' $\vert\Phi_j\rangle\equiv \overline{O}_j\vert 0\rangle$,
we have limited knowledge and control of the probe states so produced.
Judiciously chosen probe operators, constructed from
smeared fields, should excite the low-lying states of interest, with
hopefully little coupling to unwanted higher-lying states, and help
with classifying the levels extracted.  Small-$a$ classical
expansions can help to characterize the probe operators, and hence,
the states they produce.

\begin{figure}
\includegraphics[width=5.8in]{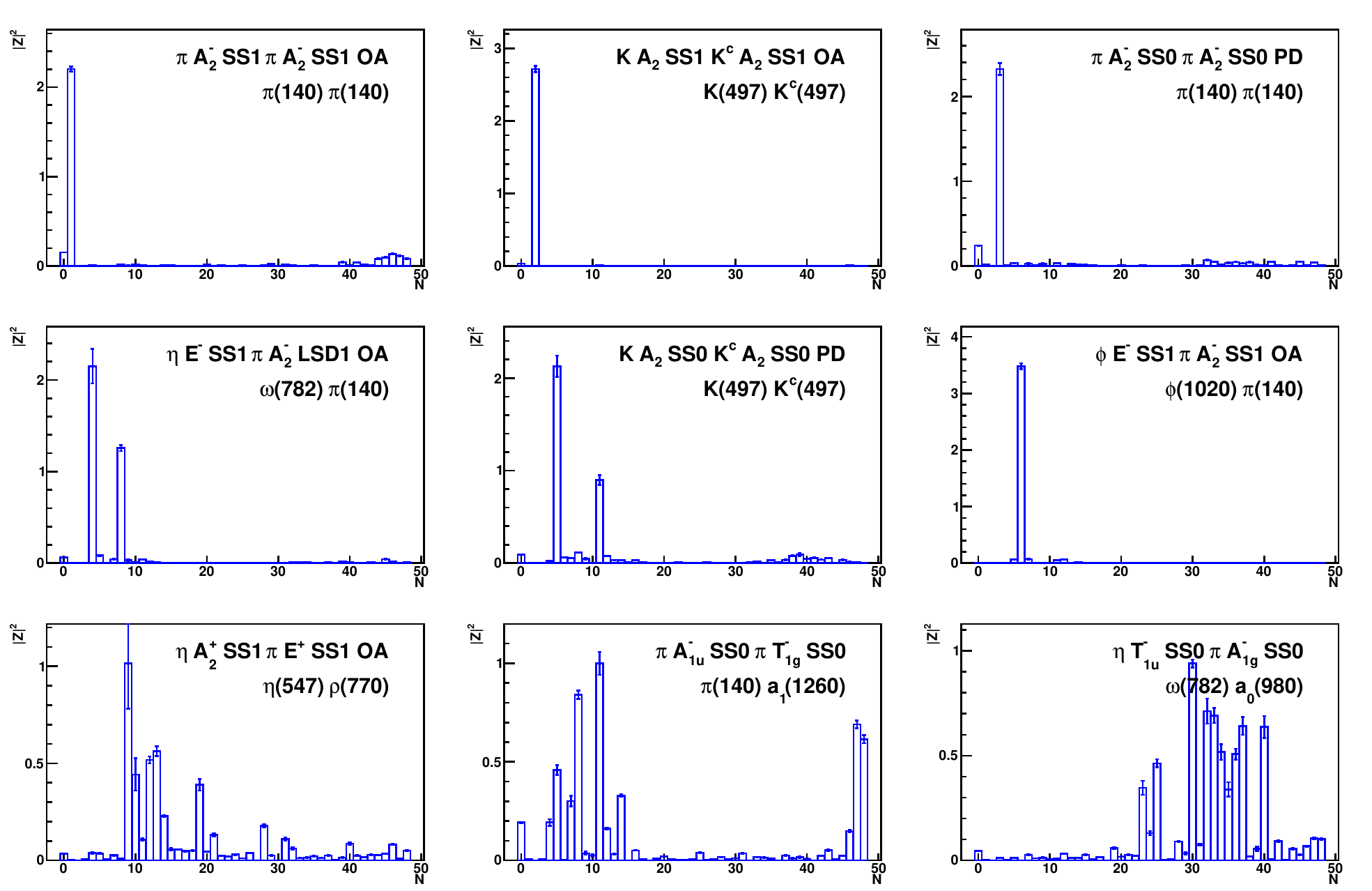}
\caption[cap6]{
Overlaps $\vert Z^{(n)}_j\vert^2$ for some selected operators $O_j$
against the eigenstates labelled by $n$. The selected operators are listed in 
the top right corner of each plot, and the physical content of the dominant
state for that operator is also listed.
The overall normalization is arbitrary in each plot.
\label{fig:Zsome}}
\end{figure}  

The overlaps $\vert Z^{(n)}_j\vert^2$ corresponding to some selected operators 
are shown in Fig.~\ref{fig:Zsome}.  In the first plot (upper left), one sees 
that this operator mainly creates level 1 (the first excited state in this
channel).  The overlaps onto all other states are very small.  This operator
is constructed from two pion operators, each having a well defined
momentum of minimal magnitude along the lattice axes; the pions have equal
and opposite momenta so that the total momentum is zero, and the momentum
directions are combined so as to produce the $T_{1u}^+$ quantum numbers. Hence, 
we can identify level 1 as dominantly a two-pion state in which the pions have
minimal relative momentum allowed for a $T_{1u}^+$ level.  An operator
expected to create a state dominated by the $\rho$ meson at rest mainly
produces level 0 (not shown), but one sees that this two-pion operator
does have some overlap onto the ground state.  We infer that level 0
is dominated by the $\rho$ at rest, but does have a small admixture of
two-pion states.  The second plot shows that its operator mainly
produces level 2 (the second excited state).  This operator is expected
to mainly create a kaon-antikaon state, so we identify this level as a
low-lying kaon-antikaon state.  For many levels, one finds
that one or a handful of probe states dominate, making classification
straightforward.  However, Fig.~\ref{fig:Zsome} also shows cases
where a given operator creates several eigenstates, making classification
of some levels problematic.

\begin{figure}
\begin{center}
\includegraphics[width=5.8in]{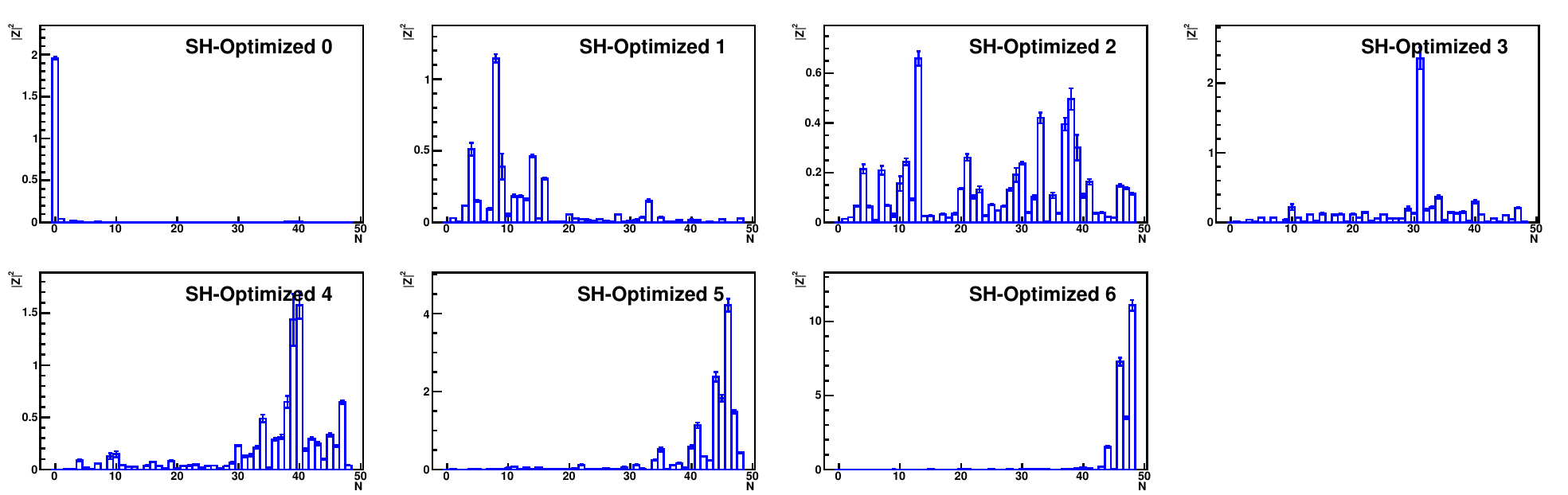}
\end{center}
\caption[cap5]{
Overlaps $\vert \widetilde{Z}^{(n)}_j\vert^2$ of ``optimized'' single-hadron
operator $\widetilde{O}_j$ against the eigenstates labelled by $n$.
The overall normalization is arbitrary in each plot.
\label{fig:Zopts}}
\end{figure}  

We will focus our efforts on level identification much more in our future
work.  For now, we mainly wish to identify the levels that dominate
the finite-volume stationary states expected to evolve into 
the single-meson resonances in infinite volume.  We view such states
as ``resonance precursor states''.  To accomplish this, we utilize
``optimized'' single-hadron operators as our probes.  We first restrict
our attention to the $12\times 12$ correlator matrix involving only
the 12 chosen single-hadron operators.  We then perform an optimization
rotation to produce so-called ``optimized'' single-hadron (SH) operators
$\widetilde{O}_j$, which are linear combinations of the 12 original
operators, determined in a manner analogous to
Eq.~(\ref{eq:rotatedcorr}).  The effective masses corresponding to these
SH-optimized operators show remarkably good plateaux at large
temporal separations, suggesting that the states created by the
single hadron operators mix very little with the states created
by the two-meson operators.  We order these SH-optimized operators according
to their effective mass plateau values, then evaluate the overlaps 
$\widetilde{Z}_j^{(n)}$ for these SH-optimized operators using
our analysis of the full $56\times 56$ correlator matrix.  The
results are shown in Fig.~\ref{fig:Zopts}.

\begin{figure}
\begin{center}
\includegraphics[trim=0mm 10mm 0mm 10mm,width=3in,clip=true]{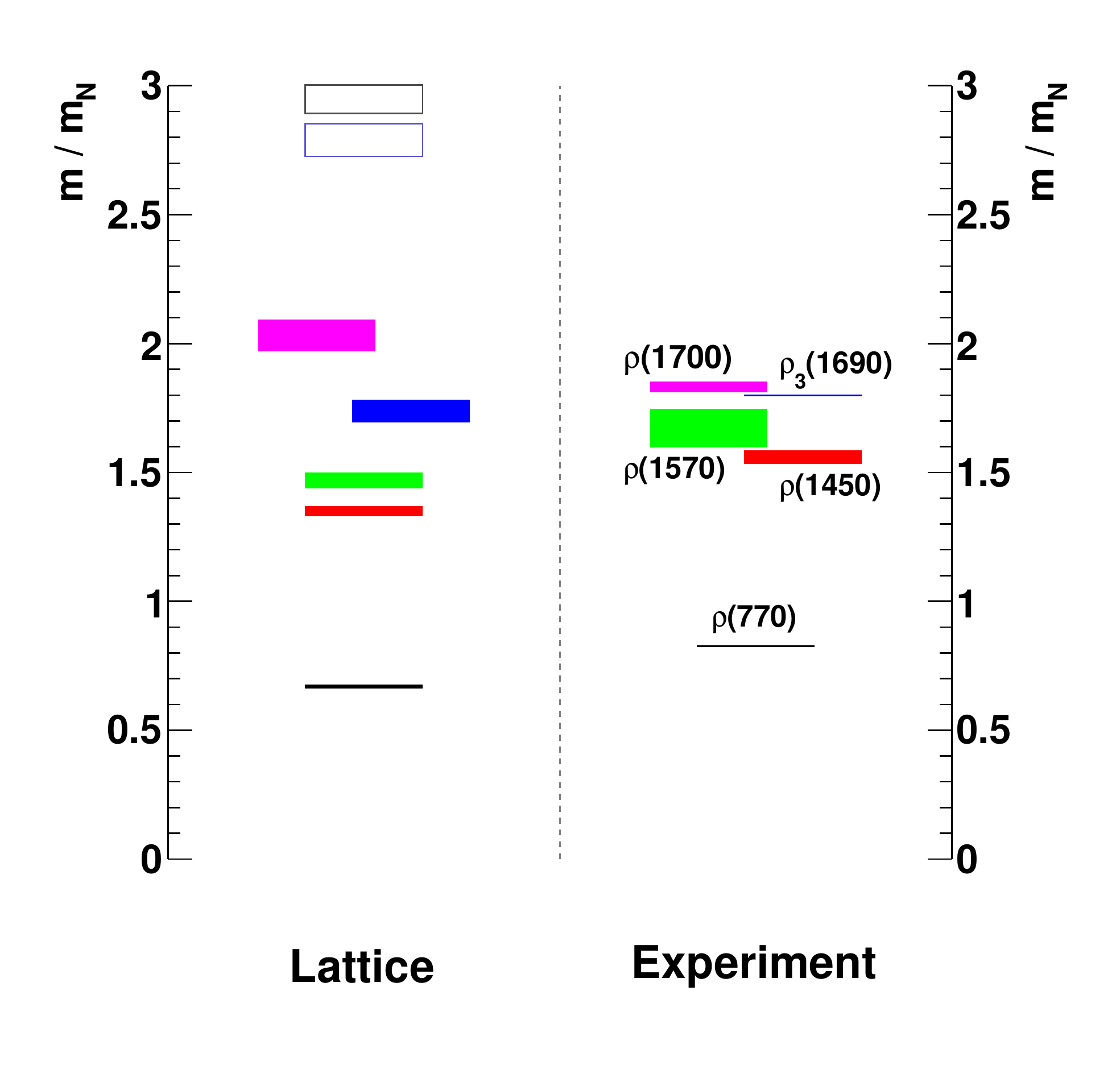}
\end{center}
\caption[cap4]{
(Left) Masses, as ratios of the nucleon mass, of the dominant
finite-volume isovector $T^+_{1u}$ stationary states expected to evolve into 
the single-meson resonances in infinite volume, computed using our
$56\times 56$ correlation matrix for the $(24^3\vert 390)$ ensemble.
The vertical thickness of each box indicates its statistical uncertainty.
The hollow boxes at the top show higher-lying states that we extract with 
less certainty due to the expected presence of lower-lying two-meson states
that have not been taken into account.
(Right) Masses of the experimentally-known spin-1 and spin-3 resonances
in this energy range.  Vertical thickness indicates experimental error.
\label{fig:boxplot}}
\end{figure}  

The first plot shows that the lowest-lying SH-optimized operator
produces level 0 and very little else.  Hence, we identify level 0
with the lowest-lying resonance precursor state, expected to be
the $\rho(770)$.  The second plot shows that this operator
produces mainly level 8, but the overlaps onto a few other states
are nonnegligible.  Hence, we identify level 8 as the dominant
state that is the precursor of the first-excited resonance in this channel,
expected to be the $\rho(1450)$.  Note that the energy of level 8
is $0.3092(33) a_t^{-1}$, which is close to the energies of the other 
levels with nonnegligible overlap, so identifying this energy value 
with the $\rho(1450)$ is reasonable.  Similarly, we identify levels 13,
31, 40, and 46 as dominantly produced by our SH-optimized
operators.  To help with identification, we have devised a few
operators whose classical small-$a$ expansions have no contribution
from spin-1; in other words, these operators should produce states
dominated by spin-3 (radiative corrections could introduce $J=1$
components, but we expect these to be suppressed).  We plan to 
use such operators to help confirm that level 31 is the precursor
state corresponding to the spin-3 $\rho_3(1690)$.  

Using the energies of levels 0, 8, 31, 40, and 46, we summarize 
our single-hadron
spectrum (the eigenstates dominated by the resonance precursor
states) in Fig.~\ref{fig:boxplot}.  This figure shows the masses
as a ratio of the nucleon mass.  Given that our pion mass is around
390~MeV and that our states are extracted in finite volume, precise 
agreement with experiment is certainly not
expected.  However, the general pattern of states is well reproduced.
We believe we have extracted all two-meson states that lie in the
range of the lowest-lying five single-hadron states.  We also find
two more higher lying states that couple mainly to single-hadron
operators (indicated by the hollow boxes), but there are two-meson
states lying below these that have not been taken into account,
so we view their extractions as particularly tentative.  Again,
we mention that three and four meson states are not taken into
account at all.

\section{Conclusion}
\label{sec:conclude}

In this talk and accompanying poster, our progress in computing
the finite-volume stationary-state energies of QCD was described.
Our first results in the zero-momentum bosonic $I=1,\ S=0,\ T_{1u}^+$ symmetry 
sector of QCD using a correlation matrix of 56 operators were presented.  
In addition to a dozen spatially-extended meson operators, an 
unprecedented number of 44 two-meson operators were used, involving a wide 
variety of light isovector, isoscalar, and strange meson operators of varying 
relative momenta.  All needed Wick contractions were efficiently evaluated using 
the stochastic LapH method.   Issues related to level identification 
were discussed.

These first results are very encouraging, and preliminary results for
the $(32^3\vert 240)$ ensemble look even more promising, but further 
work is needed to finalize the spectrum in this channel.  Of course, 
there are many other symmetry channels to investigate, which we plan
to do in the near future.
This work was supported by the U.S.~NSF
under awards PHY-0510020, PHY-0653315, PHY-0704171, PHY-0969863, and
PHY-0970137, and through TeraGrid/XSEDE resources provided by 
TACC and NICS under grant numbers TG-PHY100027 and TG-MCA075017.

\end{document}